\documentclass[superscriptaddress,twocolumn,showpacs,preprintnumbers,amsmath,amssymb,aps]{revtex4}

\usepackage{graphicx}% Include figure files
\usepackage{dcolumn}% Align table columns on decimal point
\usepackage{natbib}
\usepackage{bm}% bold math
\usepackage{color}

\begin{document}                  % DO NOT DELETE THIS LINE

\title{The commensurate phase of multiferroic HoMn$_2$O$_5$\\ studied by X-ray magnetic scattering}

%\shorttitle{The magnetism of Holmium in the multiferroic $HoMn_2O_5$}

%file:///home/laurent/Science/Papers/RMn2O5_PRB_2007/RMn2O5_Edited.tex
\author{G. Beutier}
     \email{guillaume.beutier@diamond.ac.uk}
    \affiliation{Diamond Light Source, OX11 0DE, United Kingdom}
\author{A. Bombardi}
    \affiliation{Diamond Light Source, OX11 0DE, United Kingdom}
\author{C. Vecchini}
    \affiliation{ISIS, Rutherford Appleton Laboratory - STFC, OX11 0QX, United Kingdom}
\author{P. G. Radaelli}
    \affiliation{ISIS, Rutherford Appleton Laboratory - STFC, OX11 0QX, United Kingdom}
\author{S. Park}
    \affiliation{Department of Physics and Astronomy, Rutgers University, Piscataway, New Jersey 08854, USA.}
\author{S-W. Cheong}
    \affiliation{Department of Physics and Astronomy, Rutgers University, Piscataway, New Jersey 08854, USA.}
\author{L. C. Chapon}
    \affiliation{ISIS, Rutherford Appleton Laboratory - STFC, OX11 0QX, United Kingdom}
%\keywords{multiferroic,resonant x-ray scattering}

\date{\today}

\begin{abstract}
The commensurate phase of multiferroic HoMn$_2$O$_5$ was studied by X-ray magnetic scattering, both off resonance and in resonant conditions at the Ho-$L_3$ edge. Below 40 K, magnetic ordering at the Ho sites is induced by the main Mn magnetic order parameter, and its temperature dependence is well accounted for by a simple Curie-Weiss susceptibility model.  A lattice distortion of periodicity twice that of the magnetic order is also evidenced. Azimuthal scans confirm the model of the magnetic structure recently refined from neutron diffraction data for both Mn and Ho sites, indicating that the two sublattices interact via magnetic superexchange.
\end{abstract}


\maketitle                        % DO NOT DELETE THIS LINE

\newpage

There has been very significant recent interest in magnetoelectric multiferroics, in particular magnetically frustrated transition-metal oxides such as TbMnO$_3$\cite{Kimura_Nature03}, Ni$_3$V$_2$O$_8$\cite{Lawes_PRL05}, RMn$_2$O$_5$ (R=rare-earth, Bi) \cite{Hur_Nature04,chaponPRL_Y,chaponPRL_Tb,blakePRB,vecchini,MunozPRB_Bi} and more recently RbFe(MoO$_4$)$_2$ \cite{RbFe}. In this class of materials, the onset of ferroelectricity coincides with the transition to a magnetically ordered state characterized by a complex spin configuration. While the spontaneous electric polarization is much weaker than for proper ferroelectrics, the magneto-electric coupling is strong, leading to giant effects in the physical properties upon application of an electric or a magnetic field. Some key facts about these materials have been established unambiguously: acentric magnetic ordering on the transition-metal sites is clearly the main factor in inducing ferroelectricity, and the direction of the electrical polarization can be well explained by the Landau phenomenological analysis of the magneto-electric coupling or by more general symmetry arguments \cite{radaelli-2006}.  Other aspects are much less clear: the microscopic origin of ferroelectricity is still much debated, in particular in the RMn$_2$O$_5$ and triangular-lattice systems, where more than one mechanism has been proposed. In addition, there is presently very little understanding of the role played by the magnetic rare earths. Rare earth magnetism not only leads to spectacular magneto-electric effects at low temperatures, but  can increase greatly the value of the electrical polarization even at high temperature, as recently shown for  DyMnO$_3$ \cite{Prokhnenko_PRL07}.

\indent Most $RMn_2O_5$ become ferroelectric below $\sim$ 40 K\cite{Hur_Nature04}, driven by magnetic ordering to a commensurate state with propagation vector k=(1/2,0,1/4) for R=Y,Ho,Tb \cite{chaponPRL_Tb,blakePRB} and k=(1/2,0,1/2) for Bi \cite{MunozPRB_Bi}. At around 20K, the R=Y, Ho, Tb analogues show a second magnetic transition to an incommensurate structure, with much smaller spontaneous electric polarization. Neutron diffraction studies of the ferroelectric/commensurate phase on powder\cite{chaponPRL_Y,chaponPRL_Tb,blakePRB} and more recently single crystal specimens\cite{vecchini,Kimura_JPSJ07}, have evidenced a small moment on the magnetic rare-earth sites well above the main ordering temperature of the rare earth sublattice ($\sim$ 10 K), suggesting that this moment is induced by the complex Mn magnetic ordering. For HoMn$_2$O$_5$, the magnitudes and directions of the Ho moments has been established independently by two groups \cite{vecchini,Kimura_JPSJ07}, but there cannot be absolute confidence in these results, since Ho's overall contribution to the magnetic neutron scattering intensities is very small (the Ho moments are between 0 and 1 $\mu_B$ on different lattice sites compared with an almost fully ordered Mn sublattice). More importantly, the nature of the interplay between Mn and Ho order --- in particular the induced nature of the Ho moments --- are impossible to address with neutron diffraction, since the Ho and Mn contributions cannot be separated at temperature close to T$_N$  \\
\indent In the present Communication, we report on resonant and non-resonant X-ray magnetic diffraction experiments performed on the commensurate phase of HoMn$_2$O$_5$. Unlike neutron diffraction, this technique allows one to probe the Ho and Mn ordering independently. The magnetic structure of the Ho sublattice was confirmed by fitting azimuthal scans in resonant conditions at the Ho L$_3$ edge, while off-resonance scans can be well reproduced with the overall magnetic structure determined by neutron diffraction. By studying the temperature dependence of both signals, the Ho and Mn sublattice magnetic order parameters are extracted, clearly showing that the Ho and Mn magnetic transitions are simultaneous and that the former is induced by the latter. The evolution of the Ho order parameter is well explained by a simple model accounting for the polarization by the Mn sublattice. Weak charge superlattice peaks with k$_L$=2k$_M$=(0,0,1/2) are also observed in the same temperature range; we find that the order parameter of the charge superstructure is proportional to the square of the Mn staggered magnetization, as predicted by Landau theory.\\

\indent Synchrotron X-ray experiments were performed at Diamond Light Source, on Beamline I16 - Materials \& Magnetism, dedicated to X-ray diffraction in the energy range 3-25 keV. The horizontally polarized beam delivered by a linear undulator was tuned either to the $L_3$ edge of Holmium ($\sim$8.078 keV) or at 6.4 keV below the $K$ edge of Manganese ($\sim$6.55 keV), and monochromatized by a Si $(111)$ double-crystal monochromator. The beam was focused to $120\times32\ \mu$m in the horizontal and vertical directions. A single crystal of $HoMn_2O_5$, with size approximately $1\times1\times3$ mm$^3$ and faces along the $(110)$, $(1\bar10)$ and $(001)$ crystallographic directions was used. The crystal was attached to a copper platform with QuickFill and mounted on the Kappa-geometry diffractometer of the beamline, in a Displex type cryofurnace. When necessary, a polarization analysis of the diffracted beam was performed with the (022) reflection of a copper crystal. Its efficiency at 6.4 keV was about 1/4 and the cross-talk between polarization channels was found to be negligible. The intensity was recorded either with an Avalanche Photo-Diode (APD) or with an energy-resolving Silicon-drift diode detector to reject the fluorescence. Further details of the beamline are available on \cite{i16}.\\

\begin{figure}[ht]
\includegraphics[width=\columnwidth]{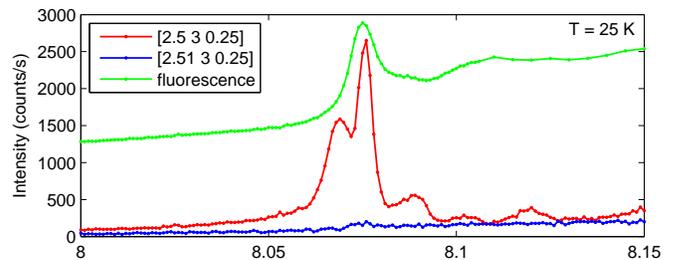}
\caption{(Color online) Energy dependence of the magnetic reflection (2.5 3 0.25) (red line) and of the fluorescence background (green line) around the Ho $L_3$ resonance energy. The quadrupole and dipole contribution at the resonance occurs at 8.068 and 8.076 keV, respectiviely. At higher energy, oscillations of the signal are observed, probably related to magnetic EXAFS, not studied here.}
\label{fig.energy}
\end{figure}

\indent All measurements were performed in the commensurate phase on two strong magnetic reflections, indexed ($\frac{5}{2}$ 3 $\frac{1}{4}$) and ($\frac{7}{2}$ 4 $\frac{1}{4}$) and on the (4 4 0.5) "charge" Bragg peak.%, which is clearly structural in origin since it is only observed in the $\sigma\rightarrow\sigma$ channel. 
The first set of measurements were conducted at 6.4 keV, i.e. at an energy far below any absorption edges of the sample. In these conditions, the magnetic scattering is not element specific and measures contributions from both Mn and Ho sites. The low fluorescence, combined with the high efficiency of the analyzer and the low noise of the APD, allowed us to measure in non-resonant conditions with good statistics (typically 1000 counts/sec) and low background (about 250 counts/sec in the $\sigma\rightarrow\sigma$ channel and nearly 0 counts/sec in the $\sigma\rightarrow\pi$ channel). No resonance was found at the $K$-edge of Manganese (not shown), confirming that its orbital moment is quenched by the crystalline electric field (at a $K$-edge, one actually measures the orbital contribution of the specific ion to the total magnetization, which is related to the spin component by the spin-orbit coupling). A second set of measurements was conducted in resonant conditions in the vicinity of the Ho $L_{3}$-edge. A strong dipole-dipole enhancement, due to the resonance of the $2p_{3/2}$ electrons with the empty $5d$ shell, was observed at 8.076 keV (Fig. \ref{fig.energy}). A quadrupolar contribution was also found at 8.068 keV. 

\begin{figure}[ht!]
\includegraphics[width=\columnwidth]{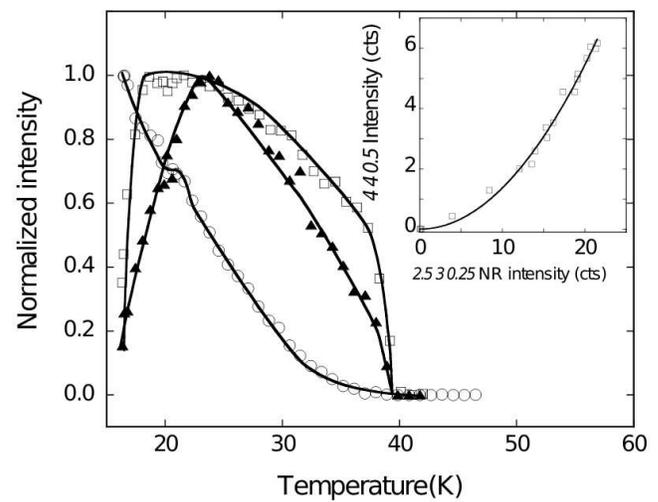}
\caption{Integrated intensities of various Bragg peaks as a function of temperature. The intensity of the magnetic reflection (2.5 3 0.25), collected on warming, is shown off resonance (energy of 6.4 keV) and at the Ho $L_3$ edge (energy of 8.076 keV) with filled triangle and open circle symbols, respectively. The superstructure charge reflection (4 4 0.5), measured off resonance at 7.96 keV, is shown as open square symbol. Lines are guide to the eyes. The inset shows the intensity of the (4 4 0.5) superstructure peak against the intensity of the (2.5 3 0.25) peak, for the temperature range 25-40 K. The line corresponds to a fit with a quadratic law.}
\label{fig.temp}
\end{figure}

\indent Figure \ref{fig.temp} shows the temperature dependence of the magnetic reflection ($\frac{5}{2}$, 3, $\frac{1}{4}$), off resonance and at the Ho-$L_3$ edge. The non-resonant measurement indicates that the transition temperature is $\sim$ 40K, in agreement with previous neutron diffraction \cite{blakePRB} and magnetization measurements. Around 18 K, the compounds undergoes a first order transition (not studied here) to an incommensurate magnetic order with propagation vector k=(0.48, 0, 0.28) \cite{blakePRB}, splitting the ($\frac{5}{2}$, 3, $\frac{1}{4}$) peak and causing the accidents observed around 18K in Fig. \ref{fig.temp}. Near the transition, the peak intensity in non-resonant conditions is essentially proportional to the square of the Mn magnetic moment, since here the Ho contribution is negligible. Its temperature dependence is consistent with the neutron data \cite{chaponPRL_Tb} and confirms that Mn ordering is the primary order parameter. In the Landau free energy expansion, the magnetic order parameter $M$ can be coupled not only to the gamma-point ferroelectric distortions, but also to an additional structural modulation ($L$) of double wave-vector, k$_L$=(0,0,0.5). Time-reversal symmetry invariance requires $L \propto M^2$ at the leading term, so we also expect the charge peak intensity to be proportional to the \emph{square} of the magnetic peak intensity.  This is illustrated in the temperature dependence of the superlattice reflection (4 4 0.5), measured off resonance at 7.96 keV and shown in Figure \ref{fig.temp}. Its integrated intensity is proportional to the square  of the integrated intensity of the ($\frac{5}{2}$, 3, $\frac{1}{4}$) magnetic reflection (inset of Figure \ref{fig.temp}). This result is similar to what was recently derived from non-resonant X-ray scattering at the $Mn$ $L_3$ edge for the analog compound TbMn$_2$O$_5$ \cite{koo_2007}. The underlying structural distortion minimizes the magnetic energy, most likely through a modulation of the $c$-axis distances (and coupling), responding to the alternation of FM and AFM configuration across the rare earth layers.\\

\begin{figure}[h!]
\includegraphics[width=\columnwidth]{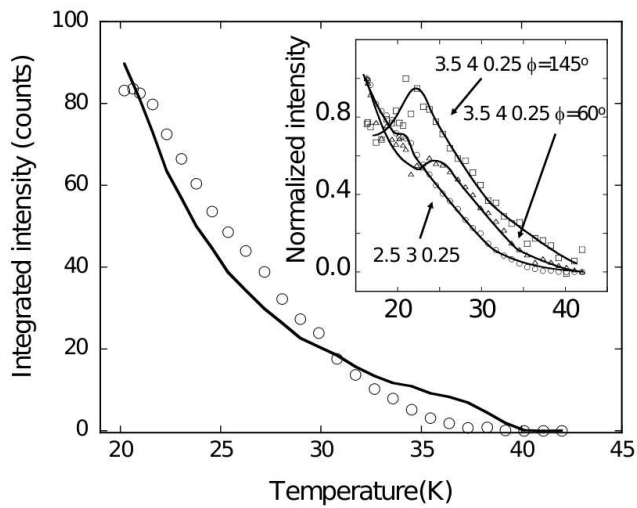}
\caption{Intensity of the resonant magnetic Bragg peak (2.5 3 0.25) as a function of temperature (open symbols). The line corresponds to a fit with a modified low-field Brillouin (Curie-Weiss) function depending on the Mn molecular field (see text for details). The inset shows the dependence for other Bragg peaks at the Ho $L_3$ edge and illustrates that the same overall behavior, with some deviations, is observed irrespective of the scattering vector or azimuth angle.}
\label{fig.tempfit}
\end{figure}
\indent In sharp contrast to the off-resonance data, the ($\frac{5}{2}$, 3, $\frac{1}{4}$) magnetic peak shows an unusual variation with temperature at the Ho-$L_3$ resonance. The peak appears below 40K, clearly showing that magnetic ordering of the Ho moments coincides with the magnetic ordering of the Mn sublattice but its intensity is marked by the absence of critical behavior at T$_N$. Irrespective of the nature of the Mn-Ho coupling, discussed later, one can postulate that the Ho ordering is induced by the molecular field arising from the Mn sublattice. This field can be taken to be directly proportional to the magnetic moment on Mn sites, as estimated from the root-squared intensity of the non-resonant peak. The resonant intensity is then simply calculated using a Curie-Weiss function diverging at $\theta$=10 K --- the ordering temperature of the Ho sublattice --- as $I_{res}=\kappa \, \frac{I_{nr}}{\left(T-\theta\right)^2}$.  There is one adjustable parameter, $\kappa$, allowing the relative intensities of the two peaks to be scaled. Fig. \ref{fig.tempfit} shows the resonant ($\frac{5}{2}$, 3, $\frac{1}{4}$) integrated intensity together with the best fit to the molecular field model. While the model reproduces relatively well the overall temperature dependence, there is clearly a discrepancy between the two curves. This is likely to be due to a relative reorientation of the two sublattices as a function of temperature, as confirmed by the fact that different resonant peaks have somewhat different temperature dependencies (inset of Fig \ref{fig.tempfit}).  \\

\begin{figure}[h!]
\includegraphics*[width=\columnwidth]{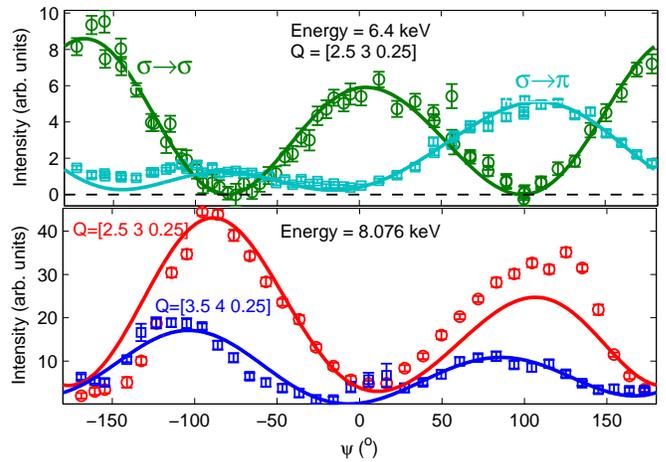}
\caption{(Color online) Azimuthal dependence of the magnetic Bragg peak intensities at 25 K. The azimuth $\psi$ value is given with respect to a reference in the [0 0 1] direction. Upper panel: (2.5 3 0.25) reflection, in non-resonant condition at 6.4 keV, in the $\sigma\rightarrow\sigma$ (green) and $\sigma\rightarrow\pi$ (light blue) channels, respectively. Lower panel: (2.5 3 0.25) (red) and (3.5 4 0.25) (blue) reflections at 8.076 keV (Ho $L_3$ edge). Symbols with associated error bars shows the experimental data points. Straight lines are calculations performed with the model of the magnetic structure from \cite{vecchini}.} %The inset shows the magnetic moments on the Ho site (red arrows) together with the calculated dipolar field from the Mn sublattice (blue arrows).}
\label{fig.azi}
\end{figure}

\indent The azimuthal dependence of the magnetic cross-section in the commensurate phase has been measured in the full angular range both off-resonance at 6.4 keV and at the Ho-$L_3$ edge (Fig. \ref{fig.azi}). For each azimuthal scan, the intensity was integrated over a rocking curve. In the non-resonant case, magnetic scattering appears in both the $\sigma\rightarrow\sigma$ and $\sigma\rightarrow\pi$ channels. At the L$_3$ edge, magnetic scattering was observed only in the $\sigma\rightarrow\pi$ channel, as expected.
The azimuthal scans are shown in Fig. \ref{fig.azi}. The scattering cross sections are proportional to the squared modulus of the scattering amplitude \cite{blume,hill}, \textit{f}, written as:
\begin{eqnarray}
    f_{nr}^{\sigma\sigma}(\mathbf{Q},\psi) &\propto& (\mathbf{k}\times\mathbf{k'}) \cdot \mathbf{S}(\mathbf{Q,\psi}) \\
    f_{nr}^{\sigma\pi}(\mathbf{Q},\psi) &\propto& (1-\mathbf{k}\cdot \mathbf{k'})\mathbf{k} \cdot \mathbf{S}(\mathbf{Q,\psi})\\
    f_{r}^{\sigma\pi}(\mathbf{Q},\psi) &\propto& \mathbf{k'} \cdot \mathbf{S_{Ho}}(\mathbf{Q,\psi})
\end{eqnarray}
where the subscript nr and r refer to the non-resonant and resonant cases and the superscript refers to the scattering channel. k and k' are unit vectors in the direction of the incident and scattered wave-vectors, respectively, and $\mathbf{S}$ are the unit-cell magnetic structure factors. Off resonance, we considered Mn and Ho spin moments with their appropriate form factors and neglected the orbital moments.
In the resonant case, only Ho sites \emph{without} the usual magnetic form factor need to be taken into account. Note that we consider only the dipole-dipole scattering, which is dominant at 8.076 keV (Fig. \ref{fig.energy}).%Off resonance, we only considered the Mn spin moment with the appropriate form factor, since the orbital contribution is negligible. \textbf{how did we treat the S and L contribution of Ho in non-resonant conditions?  It is a tiny effect of course}
The calculated intensities have been corrected for self-absorption, multiplying them by the factor:
\begin{equation}
    A(\mathbf{Q},\psi) = \frac{1}{\mu}\left[1+\frac{\sin\alpha(\mathbf{Q},\psi)}{\sin\beta(\mathbf{Q},\psi)}\right]^{-1}
\end{equation}
where $\alpha(\mathbf{Q},\psi)$  and $\beta(\mathbf{Q},\psi)$ are respectively the incident and exit angles with respect to the crystal surface (\mbox{[1 1 0]}) and the linear absorption coefficient $\mu$ does not depend on $\psi$. \\
\indent In the non-resonant case, the scattering intensities have been calculated using the model of magnetic structure for HoMn$_2$O$_5$ recently derived by single-crystal neutron diffraction at 25K \cite{vecchini}. The agreement between experimental data and calculations is excellent in both the $\sigma\rightarrow\sigma$ and $\sigma\rightarrow\pi$ channels as illustrated in Figure \ref{fig.azi}. In this model, the manganese ions form antiferromagnetic zig-zag chains along the $a$-axis, with the main components of the moment in the $ab$-plane and pointing about 14$^\circ$ with respect to the $a$-axis. A small out-of phase component along the $c$-axis gives rise to a cycloidal modulation along the Mn$^{4+}$ chains.
The azimuthal dependence at the Ho $L_3$ edge is also consistent with the neutron results, and confirms unambiguously that the Ho moments point mainly in the $ab$-plane.
This is consistent with the Ho single ion anisotropy: in fact, a point charge calculation of the Ho crystal field with the program McPhas \cite{Mcphas}, taking into account neighboring charges at distances up to 50 \AA, shows that the ground state multiplet is made of two nearly degenerated singlets. Only the matrix element $\mathbf{J_x}$ and $\mathbf{J_y}$ have non vanishing components. This is in contrast with calculations for ErMn$_2$O$_5$, which show that the ground state doublet has very strong easy-axis anisotropy, as expected from the opposite sign of the Stevens equivalent operator B$_0^2$ with respect to Ho, and also compatible with neutron diffraction study showing that the Er moments are aligned mainly along the $c$-axis \cite{Gardner_JPC1988}.
There is a residual degeneracy in the patterns of in-plane moments that can be consistent with the resonant data; however, the neutron data strongly discriminate between these alternatives. Two mechanisms, both leading to weak interactions, can account for the induced moment on the Ho site: dipole-dipole interaction on one hand and magnetic superexchange interactions between the Ho and first-neighbor Mn ions. The dipolar interaction energy, taking into account the full paramagnetic Ho moment M=10 $\mu_B$ (g$_J$=1.25,J=8), and the shortest Ho$^{3+}$-Mn$^{4+}$  distance, is 0.4K. 
%A calculation of the dipolar field induced by the ordering of the Mn ions in various points of the unit-cell is shown in the inset of Figure (Fig. \ref{fig.azi}). The Ho moment orientations, perpendicular to the dipolar field, is clearly not consistent with this mechanism. 
Calculations show that the dipolar field induced by the ordering of the Mn ions is almost orthogonal to the Ho moments, which invalidates this model.
The super-exchange interaction mediated through the Ho4(f)-O(p)-Mn4(d) orbital path is also weak given the short extension of the 4f shell. However, this mechanism is consistent with the observed structure, as already pointed out in \cite{vecchini}, since the Ho moment is oriented antiparallel to the resultant magnetic moment of first-neighbor Mn$^{4+}$ ions. \\
\indent In summary, we have studied the magnetic structure of HoMn$_2$O$_5$ in the commensurate/ferroelectric phase by magnetic X-ray scattering off resonance and at the Ho-L$_3$ resonance. Our study clearly established that magnetic ordering of the Ho sublattice exists even at temperature close to T$_N$ and is induced by the magnetic ordering of the Mn sites. The azimuthal dependence of the magnetic scattering  is in agreement with the overall magnetic structure derived by single-crystal neutron diffraction and provide an independent confirmation of the magnetic ordering on the Ho sites at 25 K. The magnetic configuration on the Ho sites is only consistent with coupling to the Mn$^{4+}$ moments through superexchange interactions, and agrees with the Ho easy-plane anisotropy derived from a point-charge model.

%\bibliographystyle{apsrev}
%\bibliography{RMn2O5_Manuscript}

\end{document}